# Structural and magnetic properties of the ordered perovskite Pb$_2$CoTeO$_6$.

**S. A. Ivanov**,
Karpov' Institute of Physical Chemistry, Moscow, Russia
**P. Nordblad, R. Mathieu**,
Department of Engineering Sciences, The Ångström laboratory,
Uppsala University, Sweden
**R. Tellgren\***,
Department of Materials Chemistry, The Ångström Laboratory, Uppsala
University, Sweden
**C. Ritter**,
Institute Laue Langevin, Grenoble, France

**Abstract**
The complex perovskite Pb$_2$CoTeO$_6$ (PCTO) has been prepared as polycrystalline powders by solid state reaction route, and the crystal structure and magnetic properties have been investigated using a combination of X-ray and neutron powder diffraction, electron microscopy, dielectric, calorimetric and magnetic measurements. It was shown that at room temperature this compound adopts a trigonal perovskite structure, space group *R-3* (a= 5.6782(1)Å, c= 13.8552(3)Å). The compound undergoes a number of temperature-induced phase transitions and adopts four different structures in the temperature range 5-500 K: monoclinic in *P2$_1$/n*(5<T<125K, tilt system ($a^+b^-b^-$)), monoclinic in *I2/m* (125<T<210 K, tilt system ($a^0b^-b^-$)) , rhombohedral in *R-3* (210<T<370K, tilt system ($a^-a^-a^-$)), and finally cubic in *Fm-3m* (above 370K without any tilting). These structural phase transitions are coupled to a change in the dielectric constant and the heat capacity around 210 and 370K. A long-range antiferromagnetically ordered state has been identified from neutron powder diffraction and magnetic studies at different temperatures. Magnetic diffraction peaks were registered below the transition at about 16 K and a possible model for the magnetic structure is proposed. Possible coexistence of long-range ordering of electrical dipoles and magnetic moments at low temperatures making PCTO a potential multiferroic candidate is discussed and compared with those of other Co-based quaternary oxides adopting the perovskite structure.

**Introduction**
Multiferroics are currently in focus of the material science and the list of multiferroics is constantly updated [1-4]. Combining ferroelectricity and magnetism in a single-phase compound would obviously be of tremendous interest not only for practical applications but also for fundamental physics [5-9]. The intrinsic ability to couple the electrical polarization to the magnetization allows an additional degree of freedom in the design of conventional devices.
Reports of large ferroelectric polarization in thin films of BiFeO3 [10] and strong magnetoelectric coupling in TbMnO3 [11] stimulated enchanced activity in the field of multiferroics. Recent observations of magnetoelectric effects in materials with complex magnetic structures, in which the spontaneous electric polarization is induced by the magnetic order, also opened additional prospects in the quest for new multiferroics [12]. From the materials chemistry point of view, ceramics with the perovskite structure present an incredible wide array of structures and phases with totally different functions, which are directly related to nature of its constituent cations, as well as on its structural features, including distortion and ordering. Perovskites have also played a leading role in multiferroic research [13].Although magnetism and ferroelectricity usually exclude each other [14], it has been known since the early 1960s that they can exist in a few materials known as multiferroic perovskites [15-19]. These rare compounds exhibit magnetic and electric orders and thus provide a unique opportunity to exploit several functions in a single material.The strategy of introducing magnetic behaviour into ferroelectric compounds was connected with an incorporation of sufficient concentration of uncompensated spins in the B-sublattice. However, limited progress has been made during the last several decades for the following reasons. First, the coupling in existing single-phase compounds was too weak to be of practical use. Secondly, there were very few compounds displaying a coexistence of dipole and spin orders and a magnetoelectric coupling. Furthermore, the Curie or Neel temperature of most of the compounds was far below room temperature. Therefore, it remains a critical need to find new single-phase materials with strong magnetoelectric coupling at room temperature for practical applications [20].

Lead-based perovskites have attracted attention because of their excellent dielectric, piezoelectric, and electrostrictive properties,which are useful in many modern applications[21]. In the same time Pb perovskites have recently been investigated in great detail because of interesting magnetic properties [12-14, 22-25]. These perovskites are good candidates to show both spin and dipole orderings. Spontaneous polarized states are expected from the distorted lead coordination. Paramagnetic ions at the B position ($Co^{2+}$ in our case) can lead to the magnetic ordering. The search of new multiferroic materials needs to clarify the role of a $6s^2$ lone pair of $Pb^{2+}$ cations, which strongly influences on the structural distortions. At the same time it was found that the presence of p-elements like tellurium could be useful in stabilising ferroelectric properties in such perovskites [17,26] if they show the required spherical symmetry and adequate ionic sizes. Although a number Te-based perovskites with simultaneous presence of Co and Te sharing their B positions have been reported [12-14,27-30], to the best of our knowledge, no detailed investigation of structure and properties of PCTO has been carried out. In order to study the potential of a given material as a possible multiferroic, the knowledge of its structural features, dielectric and magnetic properties is mandatory.

The temperature of the magnetic phase transition in complex Co-containing complex perovskites [16, 17] depends on the number of possible Co–O–Co linkages. So, one possible way of changing the temperature of a magnetic phase transition is to increase the concentration of $Co^{2+}$ cations in the lattice or the degree of its ordering, as the ordering changes the number of magnetic ions in the neighboring unit cells.

**2. Motivation**

In our current research on multiferroics we have focused the interest on the study of structure, dielectric and magnetic properties of Co-based perovskite-type compounds which have promising magnetic and electric properties. The richness of the physical properties of the cobaltites is related to the ability of Co cations to adopt not only several oxidation states but also various spin states. PCTO was initially synthesised and investigated in [28] and then forgotten for a number of years. In the earliest report [28] the results of low-frequency investigations of the dielectric properties was presented and two dielectric anomalies around 210 K and 370K were discovered. It was proposed that the first of these anomalies probably corresponds to the transition from a antiferroelectric (AFE) to a ferroelectric (FE) state, while the second one was connected with the transition from a paraelectric (PE) to a AFE phase. Later it was found, that following the results of high-frequency studies of the dielectric properties only one antiferroelectric phase transition, accompanied by anomalous dielectric properties occurs in PCTO at about 367K [31]. The structural properties of PCTO have not yet been studied in detail, and the role of the structural change on the magnetic and dielectric properties is not clear. There is a controversy concerning the crystal structure of PCTO, particularly with respect to the true symmetry group [28,32]. The symmetry of PCTO is still a matter for discussion and its crystal chemistry is much more complicated than originally assumed. Although an undistorted cubic structure (a=8.00 Å) with a complete ordering in the distribution of the Co and Te cations was reported for PCTO at room temperature [28], it has been a conflicting report [32] where PCTO described as having a tetragonal structure (s.g. I4/mmm, a=5.661(5) and c=8.004(7) Å). In the light of the above considerations, further structural investigations of PCTO are appropriate. The precise knowledge of the nuclear and magnetic structures of this perovskite has a paramount importance in the interpretation of ferroic and magnetic properties. An understanding of how the chemical environment affects the coexistence of spin and dipole ordering is the main aim of this investigation. In this paper we supply previous studies with a high-resolution NPD investigation of the thermal evolution of the crystal structure and the long-range magnetic ordering of PCTO. The study is complemented with magnetic and calorimetric measurements, which are discussed in the light of subtle structural peculiarities. Additional consideration of similarities and differences between the structures of Co-based multiferroic perovskites of varying chemical compositions may help to determine the factors responsible for obverved dielectric and magnetic properties in these materials. PCTO itself may also be regarded as a model system for studying the relationship between ferroic and magnetic properties.

## 3. Experimental

### 3.1 Sample preparation

A high quality polycrystalline sample of PCTO was prepared by a conventional solid-state sintering procedure. $PbCO_3$, $CoCO_3$ and $TeO_3$ were used as starting materials. The raw materials were weighed in appropriate proportions for the $Pb_2CoTeO_6$ formula. After that, they were wet-ball milled for 24 h, dried, and then heated at 700°C for 10 h. in an $O_2$ environment. The fired powders were ground again by wet-ball milling and pressed into a pellet and then heated up at 780°C for two days and 820°C for two days in oxygen flow with several intermediate grindings, and new pellets made after each grinding step. The samples were, finally, slowly cooled to room temperature in about 15h.

### 3.2. Chemical composition
The chemical composition of the prepared ceramic PCTO samples was analyzed by energy-dispersive spectroscopy (EDS) using a JEOL 840A scanning electron microscope and INCA 4.07 (Oxford Instruments) software. The analyses performed on several particles showed that the concentration ratios of Pb, Co and Te are the stoichiometric ones within the instrumental resolution (0.05).

### 3.3. Second harmonic generation (SHG) measurements.
The material was characterized by SHG measurements in reflection geometry, using a pulsed Nd:YAG laser ($\lambda=1.064$ μm). The SHG signal $I2\omega$ was measured from the polycrystalline sample relative to an $\alpha$-quartz standard at room temperature in the Q-switching mode with a repetition rate of 4 Hz.

### 3.4. Magnetic and dielectric measurements
The magnetization experiments were performed in a Quantum Design MPMSXL 5 T SQUID magnetometer. The magnetization (M) was recorded as a function of temperature (5-300 K) in 20 and 1000 Oe field using zero-field-cooled (ZFC) and field-cooled (FC) protocols.
Dielectric measurements of PCTO ceramic samples were carried out on a disc shaped pellet with silver paste electrodes painted onto it. The dielectric constant was measured with *LCR* (Hewlett–Packard 4294A) at 1 kHz at temperatures between 100 K and 500 K in air.

### 3.5. Specific heat measurements
Specific heat measurements were performed using a relaxation method between 2 K and 50 K on a Physical Properties Measurement System (PPMS6000) from Quantum Design Inc. Heat capacities in the temperature range 100-500 K were obtained with help of a differential scanning calorimetry technique by using a DSCQ1000 from TA Instruments with ceramic PCTO samples sealed in aluminium pans. The heating rate was maintained at 2$^{o}$C min$^{-1}$.

### 3.6. X-ray diffraction
The phase identification and purity of the powder sample was checked from X-ray powder diffraction (XRPD) patterns obtained with a D-5000 diffractometer using Cu K$_\alpha$ radiation. The ceramic sample of PCTO were crushed into powder in an agate mortar and suspended in ethanol. A Si substrate was covered with several drops of the resulting suspension, leaving randomly oriented crystallites after drying. The XRPD data for Rietveld analysis were collected at room temperature on Bruker D8 Advance diffractometer (Ge-monochromatized Cu K$_\alpha$ radiation, Bragg-Brentano geometry, DIFFRACT plus software) in the 2θ range 10-152$^{°}$ with a step size of 0.02$^{°}$ (counting time 15 s per step). The slit system was selected to ensure that the X-ray beam was completely within the sample for all 2θ angles.

To determine the correct symmetries and the associated space groups of PCTO, we examined both the type of splitting of the basic reflections of the primitive cubic perovskite and the occurrence of superstructure reflections. Howard *et al.* [33] used group-theoretical methods to list 12 distinct structures arising from the rock-salt ordering of the *B* and *B*' cations in combination with various sets of in-phase and out-of-phase tilting. It is appropriate to indicate that the in-phase octahedral tilting produces superstructure reflections with (even,odd,odd) indices, and out-of-phase tilting produces superstructure reflections with (odd,odd,odd) indices. The simultaneous occurrence of both distortions will produce some extra reflections, indexed as (even,even,odd).

### 3.7. Neutron powder diffraction

Because the neutron scattering lengths of Co and Te are different, the chemical composition of the B-site cations can be observed by neutron powder diffraction (NPD) with good precision ($b_{Co}$= 2.49, $b_{Te}$ = 5.80 fm). The neutron scattering length of oxygen is comparable to those of the heavy atoms and NPD provide accurate information on its position and stoichiometry.

The neutron diffraction experiments on PCTO samples were performed at the Institute Laue-Langevin (Grenoble, France) on the powder diffractometer D1A (wavelength 1.91 Å) in the 2θ-range 10– 156,9° with a step size of 0.1°. The powdered sample was inserted in a cylindrical vanadium container. A helium cryostat was used to collect data in the temperature range 1.5-295 K. Nuclear and magnetic refinements were performed by the Rietveld method using the FULLPROF software [34]. The diffraction peaks were described by a pseudo-Voigt profile function, with a Lorentzian contribution to the Gaussian peak shape. A peak asymmetry correction was made for angles below 35° (2θ). Background intensities were estimated by interpolating between up to 40 selected points ( low temperature NPD experimental data) or described by a polynomial with six coefficients. During the refinements the two octahedrally coordinated metal cations (Co and Te) were allowed to vary their occupation on the two possible metal sites. For low-symmetry phases the refined atomic coordinates were used to calculate the magnitude of the tilting angles, using the formula, proposed in [35,36].The analysis of the coordination polyhedra around the cations was performed using the IVTON software [37].

The magnetic structure was refined in space group P1 as an independent phase in which only $Co^{2+}$ cations were included. Several magnetic models were tried in the refinement, each employing one additional refinement parameter, corresponding to the magnitude of the magnetic moment. Each structural model was refined to convergence.The variant for which the structural refinement was stable and the R factors at minimum was chosen as the final model.The K-search program was used to determine the propagation vector [34]. We used the program BASIREPS [38] to calculate the irreducible representations for our special case. This program provides us to command lines for the magnetic refinement using the FULLPROF software [34].

### 4. Results

### 4.1 Magnetic measurements

Figure 1 shows the temperature dependence of the magnetic susceptibility recorded in magnetic fields of 20 and 1000 Oe. The high-temperature data (70 < T < 300 K) obeys a Curie-Weiss law ($\chi(T)=C/(T+\theta)$) with a value of $\theta \sim -218$ K and an effective bohrmagneton number p of 6.7 $\mu_B$, in accord with the theoretically expected for $Co^{2+}$ in a high-spin unquenched $3d^7$ configuration. The maximum of $\chi(T)$ at 19 K indicates antiferromagnetic ordering with a transition temperature of about 16 K (estimated from the maximum slope of $\chi T$ vs. T). This low value of $T_N$ compared to $|\theta|$ yields a high value of the frustration parameter f=-$\theta/T_N \sim 13.6$, indicating a large magnetic frustration. The ZFC and FC curves behave quite similarly in the whole measured temperature range, although a weak anomaly is observed in the susceptibility below T* ~ 50 K. This anomaly remains also at larger fields (1000 Oe), and does not seem associated with a magnetic transition in an impurity.

**4.2 Specific heat**

At low temperatures, a relatively sharp peak is observed at 16 K in the heat capacity curve (plotted as C/T), as seen in Figure 2. This peak marks the antiferromagnetic transition detected in the magnetic susceptibility measurements. The magnetic contribution to the specific heat and entropy is estimated using a baseline connecting the data just above the peak with the data at the lowest temperatures; yielding associated lower limit values for $\Delta C$ and $\Delta S$ of 0.93 R and 0.25 R respectively. Interestingly, the C/T curve only shows a broad shoulder above $T_N$, and in the vicinity of T*, confirming the absence of long-range magnetic order at these temperatures.
Figure 3 shows the temperature dependence of the heat capacity of pure PCTO which was measured on heating. Two anomalies around 215K and 375K were observed. This fact was consistent with the dielectric constant (Figure 4) and the remarkable thermal hysteresis in the DSC response confirmed that the phase transition around 210K is probably of first order.

**4.3 Dielectric measurements**

The temperature dependence of the dielectric constant $\varepsilon$ was measured in the temperature range 100-500K (see Figure 4). The dielectric constant curve exhibited two broadened maxima around 210K and 370K. Preliminary measurements at different frequencies reveal the absence of relaxor behaviour. The value of $\varepsilon$ increases slightly as frequency decreases but its overall shape and transition temperature remain alike.
The position of dielectric anomalies are in a good agreement with the results presented in [28,31]. An additional investigation did not show a dielectric hysteresis loop up to breakdown field values, which indicates a possible antiferroelectric nature of these anomalies.
Additional detail evaluation of dielectric characteristics of PCTO dense ceramics in broad temperature and frequency ranges are in progress.

**4.4 Structural Characterization**

According to the elemental analyses performed on 20 different crystallites, the metal compositions of PCTO is $Pb_{0.98(2)}Co_{0.50(2)}Te_{0.52(2)}$, if the sum of the cations is assumed to be 2. These values are very close to the expected ratios and permit us to conclude that the sample stoichiometry is the nominal one. The oxygen content, as determined by thermogravimetric analysis, is also in agreement with the $Pb_2CoTeO_6$ formula. The microstructure of the obtained powders, observed by scanning electron microscopy, reveals uniform and fine grain distribution. The first crystallographic characterization of PCTO compound was performed by XRPD analysis at room temperature which showed that the prepared samples formed crystals with a hexagonally-distorted perovskite structure (see Figure 3). Second harmonic generation (SHG) measurements at room temperature gave a negative result, thus testifying that at this temperature the PCTO compound probably possesses a centrosymmetric crystal structure. The sample could still be non-centrosymmetric, but at a level detectable only with a sensitivity beyond $10^{-2}$ of quartz [39].

### 4.4.1 X-ray powder diffraction

XRD pattern of PCTO at 295K shows a main set of strong peaks characteristic of the primitive perovskite structure. In addition, the presence of superstructure reflections with substantial intensity indicates an ordered arrangement of the Co and Te cations. Visual inspection also shows that the line splitting is clearly visible in some of the peaks, which suggests that the true symmetry is lower than cubic.

XRD pattern was successfully indexed assuming a hexagonal unit cell with $a = \sqrt{2}a_p$ and $c = 2\sqrt{3}a_p$ ($a_p$-parameter of primitive perovskite unit). The unit cell with lattice parameters $a = 5.6782(1)$ Å, $c = 13.8852(3)$ Å was selected, with no impurity phases being detected. The systematic absences observed in the XRD pattern are compatible with several space groups. Attempts to fit a structural model were performed only for centrosymmetric space groups. The XRD pattern of PCTO shows well-resolved splitting of the *(hhh)* peaks but not the *(h00)* reflections, consistent with s.g. *R-3* (the tilt system $a^-a^-c^-$ following a classification of [33]). There was no suggestion even at the resolution of our XRD instrument that cubic [28]) or tetragonal [32] symmetry should be invoked for our PCTO sample. The centrosymmetric group *R-3* was selected leading to the best refinement with the smallest number of refined parameters. The starting model for the crystal structure of PCTO was the set of coordinates proposed for this type of perovskite distortion in [33]. Refinements of the site occupancy factors of the cations and oxygen atoms did not reveal significant changes from full occupancy (within three standard deviations) and was therefore fixed at unity. The final Fourier difference synthesis calculation showed no significant maxima. The fit obtained is displayed in Figure 5 and the structural parameters are recorded in Table 1. The lattice parameters for PCTO are not in a reasonable agreement with those reported earlier in [28]. Considering the results of [32] where PCTO was indicated as a tetragonal, in principal several possible tetragonal structures were identified from the group theoretical analysis [35], but neither of these was appropriate since they do not allow for splitting of the *(hhh)* reflections which were registered for PCTO sample at 295K. The difference in the symmetry and metrics of the unit cell could be explained by the different sample preparation methods.

Examination of XRPD patterns at 500 K has clearly revealed a symmetry change and structural phase transition. However, since PCTO contains heavy atoms, the intensity of the superstructure reflections associated with the tilting may be too weak to be revealed by XRPD technique and NPD is more attractive in order to distinguish the lattice distortions and select the correct space group.
.
### 4.4.2 Neutron powder diffraction
We started to refine the crystal structure of PCTO sample using NPD data at 295K. To test the space group found from the refinements of the XRD data several centrosymmetric trigonal space groups were initially considered. Different models based on partial ordering between Co and Te cations in the B-sublattice were also tested. Rietveld refinements were carried out in all space groups, but a clearly superior fit was obtained using space group *R-3* with an ordered distribution of Co and Te in the B-site previously proposed from the XRPD data. The obtained atom positions are very similar, but we were able to determine more accurately the oxygen positions due to the characteristics of the neutron scattering. No vacancies were observed in the cationic or in the anionic substructures. Accordingly, the chemical composition seems to be very close to the nominal one and therefore, the cobalt oxidation state can be assumed to be two.
The atomic coordinates and other relevant parameters are tabulated in Table 1. Selected bond lengths and angles are listed in Table 2. The value of the tilting angle at 295K is about $5^o$.
NPD measurements were also performed at different temperatures between 5 and 500 K in order to check whether any phase transition occurred and in order to follow the evolution of the nuclear and magnetic structures at low temperatures. The diffraction profiles are shown in Figure 6. There is clear evidence of several phase transitions in the crystal structure between 5 K and 500 K. From inspection of these patterns and considering only those possibilities enumerated in [33], the identification of the structures becomes relatively straightforward. The structure has been taken as cubic when all reflections are single peaks, rhombohedral when the cubic *(hhh)* peaks are split but not the *(h00)* ones and monoclinic when both the *(h00)* and *(hhh)* (among others) show splitting.
In Figure 7, the thermal evolution of the cubic (444) and (400) reflections is shown. The (444) reflection is appearing as a singlet in *Fm-3m* and *I4/m* but can be split in both *I2/m* (s.g.12-2) and *R-3*. In the former case a triplet with an intensity ratio of 1:1:2 is expected, even if the monoclinic angle in near 90 this may be a doublet. In the latter case the (444) reflection should be a doublet with an intensity ratio of 3:1. The (400) reflection is unsplit in *Fm-3m* and *R-3*, but splits in *I4/m* and *I2/m*.
In the temperature range 210-350 K the rhombohedral angle is increasing with decreasing temperature from $60.08^o$ up to $60.24^o$.
At temperatures below 210K all observed superstructure reflections are indexed with the (odd,odd,odd) indices. The (444) reflection splits in a triplet and the (400) reflection splits in a doublet. This is readily explained by a monoclinic distortion related with only out-of-phase tilting where the space group *I2/m* is a logical choice [33]. The tilting angle for this phase was estimated to $7^o$.

Below 125K, besides the (odd,odd,odd) additional reflections, several very weak peaks with the (even,even,odd) indexing (most noticeably the (120) and (122) reflections ) are also visible demonstrating the presence of in-phase $BO_6$ tilting. These together, with another superstructure reflections, show PCTO to be monoclinic in space group $P2_1/n$. The low-temperature monoclinic structure at 5K has been solved and refined in the space group is $P2_1/n$ (the tilt system a$^-$ a$^-$ c+) where the combination of cation ordering and the out-of-phase tilting around the *y*-direction, and the in-phase tilting around the z-axis are existing. The magnitude of the tilt about the z axis (around 3$^o$) is to be smaller than that about the y axis (about 8$^o$) but both these tilts are significantly greater than zero. The alternate monoclinic space group *I2/m* can be discounted from this evidence

At the temperature above 210K the additional superstructure reflections arise from the in-phase tilt of the octahedra. The basic (444) reflection is a doublet with the intensity ratio of about 3:1; the basic (400) reflection does not split (Figure 7). This is characteristic of a rhombohedral symmetry. Since the observed additional reflections in PCTO correspond to the out-of-phase tilting of the octahedra, the space group *R-3* is evident. Above 350K we did not observe any peak splitting in the NPD patterns of PCTO indicating that the compound adopts the cubic space group *Fm-3m*.

Details of the structures at different temperatures, corresponding to the four distinct phases of PCTO are recorded in Table 1. The corresponding patterns have been presented in Figure 8. The most relevant bond distances and angles are given in Table 2, along with the bond-valence sums [40,41] for each atom. The observed values of the bond distances within the octahedra are in good agreement with the values predicted from ionic radii [42]: 2.89Å for Pb-O, 2.145 Å for Co-O and 1.96Å for Te-O. The bond-valence sums show that the coordination environment for these cations and oxygen are as expected (see Table 3). The high quality of the obtained fits leaves little doubt that the structures reported here are chemically reasonable.

The temperature dependence of the lattice parameters is expected to detect the occurrence of structural phase transitions and may indicate their nature. The thermal evolution of the lattice parameters of PCTO are presented in Figure 9, which shows the lattice parameters in all phases, scaled to be comparable with the cell edge of the cubic prototype perovskite. It can be seen that several phase transitions are evident. The contraction of the *a*-axis and extension of the *b* and *c*-axis was registered in both monoclinic phases during the heating. The rhombohedral distortion of the unit cell decreases with increasing temperature and around 350 K transforms into a cubic cell. The cubic phase shows the usual contraction with decreasing temperature.

For all registered phases of PCTO we have not found evidence of antisite defects, i.e. substitution Co/Te disorder. Therefore, the structure can be viewed as an ordered stacking of $CoO_6$ and $TeO_6$ octahedra sharing corners. The Pb atoms occupy the remaining free space with an irregular 12-fold coordination. For the cubic phase a large isotropic temperature factor for the Pb atoms is noticeable. This was also observed in related double perovskites [43-45] and was explained in terms of a local atomic disorder of the $Pb^{2+}$ cations. The oxygen atoms also show large temperature factors though in a smaller magnitude.

Figure 10 displays the structures of different phases of PCTO. Clearly, the low symmetry phases show tilting octahedra, coupled to displacements of the Pb atoms from the cubic special position. Such displacements could be responsible for a possible antiferroelectric arrangement. The main displacement corresponds to $Pb^{2+}$ cations is around 0.1Å. This asymmetric environment for the Pb atoms was explained in terms of the lone-electron pair of the $Pb^{2+}$ cation. Taking into account the results of performed structural analysis, there are two main contributions to the phase transition: the different tilt of Co and Te octahedra and the displacement of $Pb^{2+}$ cations.

### 4.4.3. Magnetic structure

A comparison of the NPD patterns shows that some magnetic scattering exists below T=20 K. The magnetic contribution could be evaluated from the difference of the NPD patterns of PCTO measured above and below $T_N$ = 16 K (see Figure 6). With a temperature below T=20 K several additional reflections appear in the patterns. This indicates a long-range magnetic order in a good agreement with the magnetic measurements. As discussed in relation to the magnetic properties, the magnetic susceptibility measurements indicate an initial antiferromagnetic transition at about 16 K. As the temperature decreases, the intensity of the magnetic peaks increases. For the refinement of the high resolution data with $T < 20$ K, the presence of magnetic peaks has to be taken into account. The magnetic structure and therefore the magnetic symmetry is, however, unknown and have to be determined either by trial and error or by magnetic symmetry analysis [46-48]. The magnetic Bragg peaks can be indexed using the same nuclear unit cell; the magnetic propagation vector was proposed to be $k = [0, 0, 0]$ and we can conclude that the magnetic structure is antiferromagnetic and commensurate with the nuclear lattice. The results of magnetic symmetry analysis for the determination of the allowed irreducible representations of $k = [0, 0, 0]$ in space group $P2_1/n$ is presented in Table 4. The different magnetic arrangements were investigated, checking all the proposed basis functions. For each combination of basis function, we have carried out a systematic comparison between the observed and calculated neutron diffraction patterns at 5 K.

After checking the different solutions given in Table 4 the best agreement with the experimental data is obtained for the solution corresponding to $\Gamma_1$ with ***$m_{1X}$=-$m_{2X}$*** and ***$m_{1z}$=-$m_{2z}$*** (***$m_{1y}$ =$m_{2y}$ =0***). As shown in Table 1 at T =5 K the modulus of the magnetic moment for the Co atoms is 2.43(3) μB, which is close to the theoretical value for high spin $Co^{2+}$ ion ($d^7$) (3 μB). The reduction of the ordered magnetic moment compared with the pure ionic configuration can be due to the covalence effects.

A view of the magnetic structure associated with $\Gamma_1$ is presented in Figure 11 where the largest component of the magnetic moments is lying along the *z* direction. The coupling between the Co magnetic moments within the *z* =0 and *z* =1/2 planes is ferromagnetic and these planes are antiferromagnetically coupled.

The selected solution of the magnetic structure belongs to the irreducible representation $\Gamma_1$, for which a non-zero ferromagnetic component along the y-direction is allowed. Therefore, a small canting of the Co magnetic moments out of the (*x,z*) plane is possible. This is in agreement with the weak ferromagnetism observed in the magnetic measurements. Available experimental NPD data could not be satisfactorily fitted for any models with a change in the spin orientation of the moments out of (*x,z*) plane.
All attempts to refine some FM magnetic component along y-direction using NPD data at T<50K were unstable and oscillatory giving the value of *$m_y$* around 0.4(0.3) μB.

## 5. Discussion

The first remark to be made is that there are two distinct octahedra, $CoO_6$ and $TeO_6$, and therefore two different tilt angles. Both tilt angles, however, are determined by the movement of the same O atom, and this implies that the tilt angles in these octahedra are not independent, but related in the inverse ratio of the Co-O and Te-O bond lengths. To the extent that these bond lengths are constant, the tilt angles will be in a fixed ratio. The second remark concerns the possibility of obtaining more than one estimate for the tilt angle of a particular octahedron. For the monoclinic structure in *I2/m*, for example, the tilt angle can be estimated from either the position of the apical oxygen $O_1$ or the equatorial oxygen $O_2$. The estimates obtained are very similar and their average is presented here. The nature of the phase transitions in PCTO can be investigated by taking the tilt angle(s) to represent the order parameter(s) in the different phases.

The refined atomic coordinates and bond distances for PCTO (see Tables 1 and 2) confirm the basic structural features of the proposed structures (Figure 10).

In principle, the monoclinic β is not so far from 90° for the low temperature phases of PCTO and the orthorhombic description of these structures is possible. Such small deviations from 90° are difficult to observe unless one is working with (*a*) very high resolution data and (*b*) samples that show very little intrinsic peak broadening.

In fact, among the space groups derived from the octahedral tilting in ordered double perovskites, only *Pnnn* belongs to the orthorhombic system [33]. Nevertheless, this space group does not allow the *(hhh)* reflections to split, that was observed for PCTO.

A search of the ICSD (Inorganic Crystal Structure Database) for double perovskites with *P*2$_1$/*n* symmetry reveals that in general the values of β are somewhat larger (90.1-90.3°) in compounds where at least one of the octahedral cations has a fairly large radius (such as a rare-earth or an alkaline-earth cation). These types of cations are more prone to tolerate small distortions of their immediate octahedral environment than smaller more highly charged cations. Hence, it is reasonable to expect that such compositions will have a more distinct monoclinic unit cell, rather than the pseudo-orthorhombic cell.

The low-temperature susceptibilities are characterized by two clear maxima in both ZFC and FC curves and a slight divergence between both graphs, indicating a possible existence of weak ferromagnetic interactions or spin-glass state. When a long range AFM ordering occurs, the dominant magnetic interaction can be either between nearest-neighbor (NN) $Co^{2+}$ cations (π-orbital overlap), which are around 5.7Å apart, or between next-nearest-neighbor (NNN) $Co^{2+}$ cations (σ-orbital overlap) (about 8Å apart) (see [49]). The proposed model of the magnetic structure of PCTO indicates that AFM interactions between nearest Co neighbors along the shorter (but not linear) Co-O-O-Co partway is stronger than NNN AFM interactions over long distances where the possible Co-O-Te superexchange pathways are almost linear (the Co-O-Te angles at 5K are between 165.8 and 169.7 °). If NNN interactions can not be neglected, it can create some frustration of the AFM structure of PCTO displaying a canting of the AFM aligned spins or spin-glass state, as discussed for measured magnetic behavior. The thermal evolution of a small proportion of Co moments which are not strongly AFM coupled can be a possible reason for the anomaly registered around 50K.

In order to get some insight into the cation distribution, we carried out bond-valence sum calculations according to Brown´s model [40, 41] which gives a relationship between the formal valence of a bond and the corresponding bond lengths. In non-distorted structures, the bond valence sum rule states that the valence of the cation ($V_i$) is equal to the sum of the bond valences ($v_{ij}$) around this cation. From individual cation-anion distances the valences of cations were calculated. The Pb, Co and Te cations exhibit the valences +1.96(1), 2.02(1) and +5.91(1), respectively (see Table 3), which slightly differ from the expected valences of these cations in this compound.

Two major factors (the difference in ionic charge and in ionic size) will determine the ordered arrangement of different cations at the B-sites in the perovskite structure [17]. It has been shown [3, 17-19] that the magnetic properties of complex perovskites strongly depend on the B-site ordering. The ionic radii of $Co^{2+}$ and $Te^{6+}$ are not so similar and this combination of cations in general favour an ordered arrangement.

It can be noted that other double perovskites $Pb_2CoMO_6$ (M=W, Mo, Re) have been studied[50-55], These compounds also order antiferromagnetically at low temperatures, all with somewhat different but still low values of $T_N$ (7 K for $Pb_2CoWO_6$ [50], 11 K for $Pb_2CoMoO_6$[54] and 32 K for $Pb_2CoReO_6$ [55]). It is important to understand the possible influence of the M cations on this difference. The difference in size of different $M^{6+}$ cations are not large enough ( 0.55Å for Re, 0.56Å for Te, 0.59 Å for Mo and 0.6 Å for Mo) to quantitative explain the different values of $T_N$, but as a tendency, the value of $T_N$ decreases with increasing $M^{+6}$ cation size. In the same time, when the M cation is an element that is stable in both the +5 and +6 oxidation states (Mo and Re), obtained phases can be ordered and disordered depending on the preparation conditions[54,55] that leads to different values of $T_N$. In the case of $Pb_2CoReO_6$ an additional magnetic contribution can be connected with the Re sublattice.

It is important to indicate that in $Pb_2CoWO_6$ a sequence of four phases was found and suggested to be either FE or AFE below 293K [51]. $Sr_2CoTeO_6$ was also suggested to undergo a phase transition from s.g.*$P2_1/n$* to *$I2/m$* to *$Fm-3m$* at 373K and 773 K [56]. If we consider other series of materials, namely $A_2CoTeO_6$ (A= Ba, Pb, Sr, Ca) (see Table 5 ) it is clear that in the case of monoclinically distorted phases (for Sr and Ca) $T_N$ slightly increases with an increase of the $A^{2+}$ cation size [27]. In the same time a magnetic frustration parameter (f), as an important factor in AFM double perovskites, was found to increase with an increasing A-cation size and led to remarkable reductions in $T_N$ and the saturated magnetic moment of the Co cations in (La,A)CoNbO_6 (A=Ba,Sr,Ca)[57 ]. The opposite tendency for $A_2CoTeO_6$ series may be related with a decrease of the covalency effect from Pb to Ca, which weakens the superexchange interactions or NN decreasing (NNN increasing) with an appearance of partial order. The number of the $Co^{2+}$-O-$Co^{2+}$ pairs determining $T_N$ does not change in $A_2Co^{2+}Te^{6+}O_6$ when varying the A-type of cations (Cd, Ca, Sr, Pb) but the experimentally obtained values of $T_N$ change. In principle, if can be some indirect changes in the oxygen-related superexchange via the change in the excitation energy between O and Co electron energy levels, as well as because of possible change in the hopping integral between these states via a possible change in the degree of the Co and Te order, the change in the lattice parameter or a change in morphology

(decrease a grain size). Thus, the change of $T_N$ does not have a simple explanation. The possible leading role of Pb in both the magnetic and ferroelectric coupling in perovskites was discussed in [58]. The ferroelectric coupling in the Pb- containing perovskites is usually enhanced owing to the strong ferroelectric activity of the Pb cation with a lone-electron pair. The magnetic coupling can be enhanced either because of the involvement of the Pb ions into the superexchange or because of the influence of the Pb-related polarization onto the magnetic properties. These factors do not exist in lead-free perovskites.

In analogy with other Co-based perovskites [17-19], however, we may assume that $T_N$ is strongly related with the value of the Co-Co distance, the Co-O-Co angle and with the type of lattice distortions. At the same time it is clear that the degree of crystallographic order/disorder at Co-sites (so called anti-site disorder) and /or oxygen deficiency should vary with the type of M cations

All members of the series $A_2CoTeO_6$ (A=Ba, Pb, Sr, Ca, Cd) also belong to the family of double perovskites with a different structural distorsion (tolerance factor) and ordering of B-type cations [17,59, 60]. It has earlier been mentioned in [28-30] that the temperature of FE or AFE phase transition $T_C$ for $A_2CoTeO_6$ perovskites is increasing with decreasing size of the A-type cations. As the effective size of the A-type cation decreases from Ba to Cd, the size of the A cation is too small for the 12-fold site within a $BO_6$ octahedral framework. $BO_6$ octahedra tilt to optimize the distorted A-O distances which eventually results in a decrease in its coordination number. In the case A=Pb, Sr, Ca, Cd this tilting does not disrupt the corner-sharing connectivity present in the ideal cubic perovskite. But in the case of Ba a tolerance factor larger than unity implies that Ba cations are too large to be accommodated in the A-site and a hexagonal perovskite lattice is formed.

It is well-known [61] that there are only three structural degrees of freedom, which may be responsible for lattice distortion in complex perovskites: a) cation shifts, b) distortions of the oxygen polyhedra coordinating the cations in the A and B-sublattices and c) tilting of the oxygen octahedra. If one takes into account the values of the polyhedra distortions (see Table 3), the ferroic properties of PCTO may be connected with the Pb sublattice, where the off-center cation displacement from the polyhedral centers is significant. At room temperature it was found, using a method proposed in [37], that the volume of the Pb polyhedra is 53.6 Å$^3$. It is worth to notice that the volume of the Co and Te polyhedra are not identical (12.2(1) and 9.5(1) Å$^3$, respectively). At the same time the off-center displacements of cations are zero for Co and Te following the symmetry restrictions, while these values for the Pb cations are considerable and increase with decreasing temperature.

## 5. Concluding remarks

Pure $Pb_2CoTeO_6$ perovskite was synthesised by conventional ceramic procedures and structurally characterised applying the Rietveld analysis of XRPD and NPD data at different temperatures. It belongs to the family of double perovskites with an ordered arrangement of $Co^{2+}$ and $Te^{6+}$ cations in the B-sublattice. The most characteristic feature of the PCTO structure, aside from the B-site cation ordering, is the tilting of the Co and Te octahedral and a displacement of the Pb cations. The precise metal-oxygen distances derived from NPD data made it possible to calculate the valences of the cations and the distortion of their polyhedra. The compound undergoes a number of temperature-induced phase transitions and adopts four different structures in the temperature range 5-500 K: monoclinic in *P2$_1$/n*(5<T<125K), monoclinic in *I2/m* (125<T<210 K) , rhombohedral in *R-3* (210<T<370K), and finally cubic in *Fm3m* above 370K. This phase sequence is a logical one expected from the group-theoretical analysis [33] and coupled to a change in the dielectric constant and a heat capacity.

A model for the antiferromagnetic structure with the Co moments was proposed. Structural and magnetic features of PCTO are considered and compared with those of other quaternary complex oxides.

**Acknowledgements**

Financial support of this research from the Royal Swedish Academy of Sciences, the Swedish Research Council (VR) and the Russian Foundation for Basic Research is gratefully acknowledged.

Table 1 Summary of the results of the structural refinements of the Pb$_2$CoTeO$_6$ sample using XRD and NPD data.

| Experiment | XRD | NPD | NPD | NPD | NPD |
|---|---|---|---|---|---|
| T,K | 295 | 5 | 175 | 295 | 450 |
| a[Å] | 5.6742(4) | 5.7155(1) | 5.7034(1) | 5.6783(1) | 8.0333(2) |
| b[Å] |  | 5.6420(1) | 5.6548(1) |  |  |
| c[Å] | 13.8519(5) | 7.9346(2) | 7.9696(2) | 13.8552(2) |  |
| α[deg.] | 90 | 90 | 90 | 90 | 90 |
| β [deg.] | 90 | 90.12(1) | 90.07(1) | 90 | 90 |
| γ [deg.] | 120 | 90 | 90 | 120 | 90 |
| s.g. | R-3 | P2$_1$/n | I2/m | R-3 | Fm-3m |
| **Pb** |  |  |  |  |  |
| x |  | -0.0027(3) | 0.5018(3) | 0 | 0.25 |
| y |  | 0.5064(5) | 0 | 0 | 0.25 |
| z | 0.2526(8) | 0.2536(4) | 0.2476(4) | 0.2515(3) | 0.25 |
| B[Å]$^2$ | 1.67(6) | 0.68(2) | 0.71(3) | 1.35(3) | 1.81(4) |
| **Co** |  |  |  |  |  |
| x | 0 | 0 | 0 | 0 | 0 |
| y | 0 | 0 | 0 | 0 | 0 |
| z | 0 | 0 | 0 | 0 | 0 |
| B[Å]$^2$ | 0.64(5) | 0.36(3) | 0.41(3) | 0.55(4) | 0.57(3) |
| **Te** |  |  |  |  |  |
| x | 0 | 0 | 0 | 0.5 | 0.5 |
| y | 0 | 0 | 0 | 0.5 | 0.5 |
| z | 0.5 | 0 | 0.5 | 0.5 | 0.5 |
| B[Å]$^2$ | 0.49(2) | 0.36(3) | 0.41(3) | 0.55(4) | 0.46(3) |
| **O$_1$** |  |  |  |  |  |
| x | 0.3291(5) | 0.0463(5) | 0.0414(2) | 0.3202(2) | 0.2611(1) |
| y | 0.1308(6) | -0.010(6) | 0.2612(3) | 0.1314(3) | 0 |
| z | 0.4183(5) | 0.2598(7) | -0.0228(2) | 0.4204(2) | 0 |
| B[Å]$^2$ | 1.56(4) | 0.79(3) | 1.21(3) | 1.48(3) | 1.87(3) |
| **O$_2$** |  |  |  |  |  |
| x |  | 0.2546(6) |  |  |  |
| y |  | 0.2729(7) |  |  |  |
| z |  | -0.0263(6) |  |  |  |
| B[Å]$^2$ |  | 0.89(3) |  |  |  |
| **O$_3$** |  |  |  |  |  |
| x |  | 0.2764(8) |  |  |  |
| y |  | 0.7515(9) |  |  |  |
| z |  | -0.0245(8) |  |  |  |
| B[Å]$^2$ |  | 0.96(3) |  |  |  |
| R$_p$,% | 5.68 | 4.27 | 4.21 | 4.16 | 3.22 |
| R$_{wp}$,% | 7.84 | 5.58 | 5.31 | 5.29 | 4.52 |
| $R_B$,% | 6.19 | 4.21 | 3.46 | 3.91 | 2.83 |
| $R_{mag}$,% | - | 9.68 | | | |
| $\mu_X$ ($\mu_B$)Co | - | 0.55(9) | | | |
| $\mu_Y$ ($\mu_B$)Co | | 0 | | | |
| $\mu_Z$ ($\mu_B$)Co | | 2.29(5) | | | |
| $\mu$ ($\mu_B$)Co | | 2.35(6) | | | |
| $\chi^2$ | 2.31 | 2.14 | 2.16 | 2.25 | 2.28 |

Table 2. Selected bond distances from neutron powder refinements of the $Pb_2CoTeO_6$ sample at various temperatures

| Bonds, Å | | 5K | | 175K | 295K | 450K |
|---|---|---|---|---|---|---|
| Pb | $O_1$ | 2.614(2) | $O_1$ | 2.623(1) | 2.830(1) x3 | |
| | $O_1$ | 2.755(1) | $O_1$ | 2.838(1) x2 | 2.845(1) x3 | |
| | $O_1$ | 2.915(1) | $O_1$ | 3.084(1) | 2.678(1) x3 | |
| | $O_1$ | 3.109(2) | $O_2$ | 2.688(1) x2 | 3.004(1) x3 | |
| | $O_2$ | 2.623(1) | $O_2$ | 2.716(1) x2 | | |
| | $O_2$ | 2.730(1) | $O_2$ | 2.948(1) x2 | | |
| | $O_2$ | 2.971(2) | $O_2$ | 2.979(1) x2 | | 2.841(2) x12 |
| | $O_2$ | 2.993(2) | | | | |
| | $O_3$ | 2.564(1) | | | | |
| | $O_3$ | 2.804(1) | | | | |
| | $O_3$ | 2.890(2) | | | | |
| | $O_3$ | 3.054(2) | | | | |
| Co | $O_1$ 2.079(1) x2 | | $O_1$ | 2.089(1) x2 | | |
| | $O_2$ 2.122(1) x2 | | $O_2$ | 2.107(1) x4 | 2.094(1) x6 | 2.085(1) x6 |
| | $O_3$ 2.129(1) x2 | | | | | |
| Te | $O_1$ 1.911(1) x2 | | $O_1$ | 1.923(1) x2 | | |
| | $O_2$ 1.919(1) x2 | | $O_2$ | 1.925(1) x4 | 1.929(1) x6 | 1.931(1)x6 |
| | $O_3$ 1.925(1) x2 | | | | | |
| Co-$O_1$-Te° | 165.8(3) | | | 166.5(2) | 170.7(2) | 180 |
| Co-$O_2$-Te° | 167.3(3) | | | 169.6(2) | | 180 |
| Co-$O_3$-Te° | 169.7(3) | | | | | 180 |

Table 3 Polyhedral analysis for different phases of $Pb_2CoTeO_6$ sample (cn - coordination number, x – shift from centroid, ξ- average bond distance with a standard deviation,

V- polyhedral volume, ω- polyhedral volume distortion.

T=450K s.g.Fm3m

| Cation | cn | x(Å) | ξ (Å) | V(Å$^3$) | ω | Valence |
|---|---|---|---|---|---|---|
| Pb$_1$ | 12 | 0 | 2.841 | 53.9(1) | 0 | 1.89 |
| Co$_1$ | 6 | 0 | 2.102 | 12.4(1) | 0 | 2.14 |
| Te$_1$ | 6 | 0 | 1.929 | 9.6(1) | 0 | 5.93 |

T=295K s.g.R-3

| Cation | cn | x(Å) | ξ (Å) | V(Å$^3$) | ω | Valence |
|---|---|---|---|---|---|---|
| Pb$_1$ | 12 | 0.016 | 2.839+/-0.120 | 53.6(1) | 0.077 | 1.96 |
| Co$_1$ | 6 | 0 | 2.094 | 12.2(1) | 0 | 2.02 |
| Te$_1$ | 6 | 0 | 1.926 | 9.5(1) | 0 | 5.91 |

T=175K s.g.I2/m

| Cation | cn | x(Å) | ξ (Å) | V(Å$^3$) | ω | Valence |
|---|---|---|---|---|---|---|
| Pb$_1$ | 12 | 0.049 | 2.837+/-0.149 | 53.3(1) | 0.080 | 1.91 |
| Co$_1$ | 6 | 0 | 2.093+/-0.009 | 12.1(1) | 0.003 | 2.10 |
| Te$_1$ | 6 | 0 | 1.925+/-0.001 | 9.4(1) | 0.002 | 5.89 |

T=5K s.g.P2$_1$/n

| Cation | cn | x(Å) | ξ (Å) | V(Å$^3$) | ω | Valence |
|---|---|---|---|---|---|---|
| Pb$_1$ | 12 | 0.075 | 2.835+/-0.181 | 53.4(1) | 0.081 | 2.07 |
| Co$_1$ | 6 | 0 | 2.091+/-0.025 | 12.2(1) | 0.002 | 2.32 |
| Te$_1$ | 6 | 0 | 1.918+/-0.006 | 9.4(1) | 0.002 | 5.48 |

Table 4 Irreducible representations of the small group G$_K$ obtained from the space group P2$_1$/n for k=(0 0 0 ) and the corresponding basis vectors. The symmetry elements are written according to [48].

| Repres. irred. | h$_1$ | h$_3$ | h$_{25}$ | h$_{27}$ | Basis vectors | |
|---|---|---|---|---|---|---|
| | | | | | Co$_1$(0,0,0) | Co$_2$(1/2,1/2,1/2) |
| Γ$_1$ | 1 | 1 | 1 | 1 | [ 1 1 1] | [ -1 1 -1] |
| Γ$_2$ | 1 | 1 | -1 | -1 | - | - |
| Γ$_3$ | 1 | -1 | 1 | -1 | [ 1 1 1] | [ 1 -1 1] |
| Γ$_4$ | 1 | -1 | -1 | 1 | - | - |

Table 5 Crystallographic and magnetic data for A$_2$CoTeO$_6$ perovskites (AS-antisite disorder) at 295K

| M$^{+2}$ | Ca[27] | Sr[27] | Pb | Ba[60] |
|---|---|---|---|---|
| r(Å) | 1.34 | 1.44 | 1.49 | 1.61 |
| a(Å) | 5.4569(1) | 5.6417(1) | 5.6782(1) | 5.7996(1) |
| b(Å) | 5.5904 | 5.6063(1) | | |
| c(Å) | 7.7399 | 7.9239(1) | 13.8552(3) | 14.2658(3) |
| β(deg) | 90.24 | 90.12 | | - |
| V(Å$^3$) | 236.1 | 250.6 | 386.9 | 415.5 |
| s.g. | P2$_1$/n | P2$_1$/n | R-3 | P-3m |
| AS | No | No | No | No |
| T$_N$ | 10 | 15 | 16 | 15 |
| property | AFM | AFM | AFM | AFM |
| <A-O>(Å) | 2.794(3) | 2.815(10) | 2.839(6) | 2.918(6) |
| <Co-O>(Å) | 2.110(2) | 2.067(7) | 2.094(3) | 2.149(5) |
| <Te-O>(Å) | 1.930(2) | 1.937(7) | 1.929(4) | 1.921(6) |
| Co-O-Te (deg) | 148.1-151.5 | 163.5-169.3 | 170.7 | 177.3-178.1 |

**Figure captions**

Figure 1 Temperature dependence of the susceptibility of $Pb_2CoTeO_6$. The main frame shows low temperature results in a field of 20 Oe and the inset a wider temperature range in 1000 Oe.

Figure 2 The specific heat ($C_p/T$) vs. temperature of $Pb_2CoTeO_6$ .

Figure 3 The specific heat ($C_p/T$) vs. temperature of $Pb_2CoTeO_6$ (100-500K)

Figure 4 Dielectric constant curve of Pb2CoTeO6 ceramics at 1 kHz.

Figure 5 The observed(full dots), calculated and difference plots(solid lines) for the fit to the XRPD pattern of $Pb_2CoTeO_6$ after Rietveld refinement of the nuclear structure at 295K. The vertical marks in the middle show positions calculated for Bragg reflections. The lower trace is a plot of the difference between calculated and observed intensities.

Figure 6 Temperature evolution of NPD patterns of $Pb_2CoTeO_6$ (the magnetic reflections are indicated by an arrow).

Figure 7 Portions of NPD patterns of $Pb_2CoTeO_6$ showing the evolution of several reflections as a function of temperature.

Figure 8 The observed(full dots), calculated and difference plots(solid lines) for the fit to the NPD patterns of $Pb_2CoTeO_6$ after Rietveld refinement of the nuclear and magnetic structure at different temperatures: 450K(a), 295K (b), 175K(c) and 5K(d). In (d), the nuclear reflection positions are shown as upper vertical marks and magnetic ones are shown as lower ones.

Figure 9 Thermal dependence of lattice parameters for $Pb_2CoTeO_6$ in the temperature range 5-500 K.

Figure 10 The structural features of four different phases of $Pb_2CoTeO_6$ shown in equivalent projections along the *y*-axis.

Figure 11 The schematic representation of   magnetic structure of $Pb_2CoTeO_6$. Diamagnetic ions are omitted.

Figure 12 Variation of the ferroic phase transition temperatures ($T_C$) for perovskite compounds of $A_2Co^{2+}TeO_6$ ($A^{2+}$=Ba, Sr, Pb, Ca,Cd) versus the $A^{2+}$-cation radius..

Author's addresses

***Sergey A. Ivanov***

Department of Inorganic Materials, Karpov' Institute of Physical Chemistry, Vorontsovo pole,10 105064 Moscow K-64, Russia
**E-mail**: ivan@cc.nifhi.ac.ru

***Per Nordblad, Roland Mathieu,***

Department of Engineering Sciences, The Ångstrom Laboratory , Uppsala University, Box 534, 751 21 Uppsala, Sweden
**E-mail**: per.nordblad@angstrom.uu.se

roland.mathieu@angstrom.uu.se

***Roland Tellgren***

Department of Inorganic Chemistry, The Ångstrom Laboratory, Box 538, University of Uppsala, SE-751 21, Uppsala , Sweden
**E-mail**: roland.tellgren@mkem.uu.se

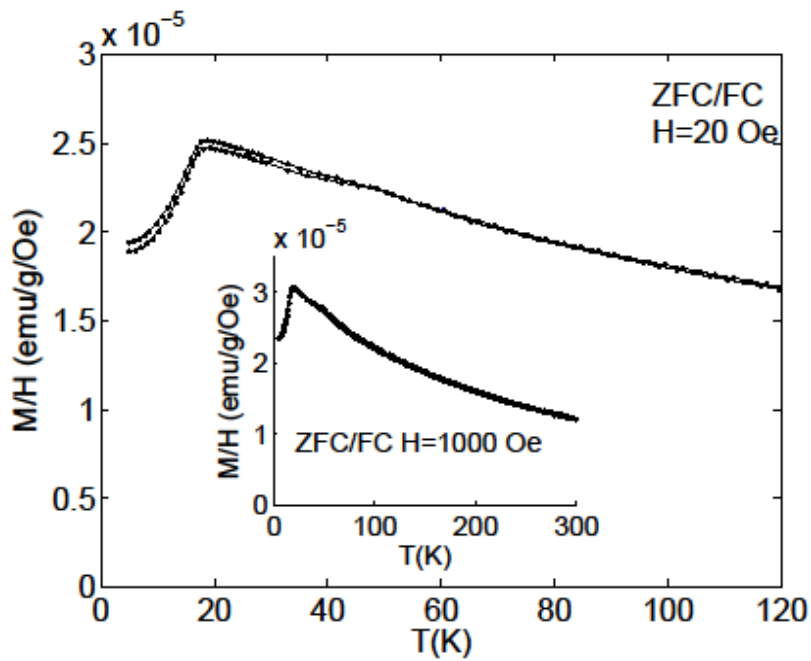

Figure 1

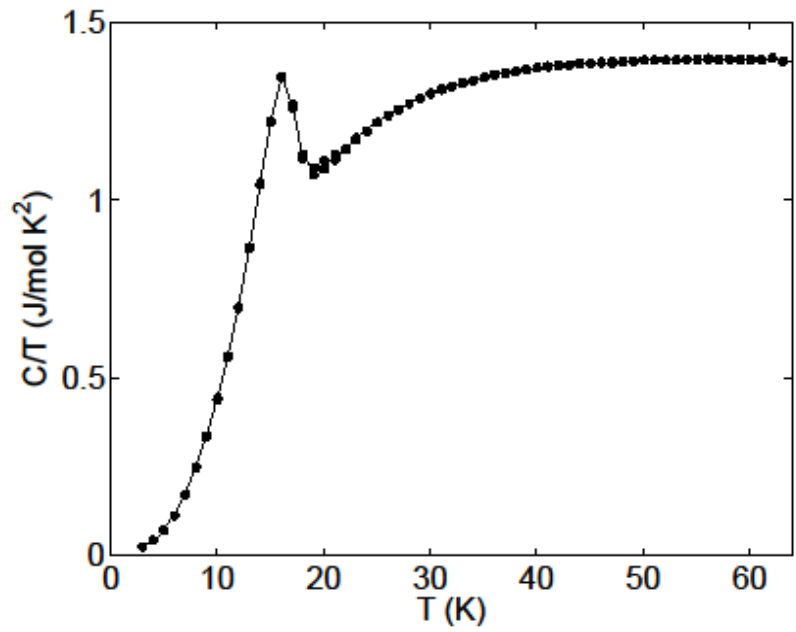

Figure 2

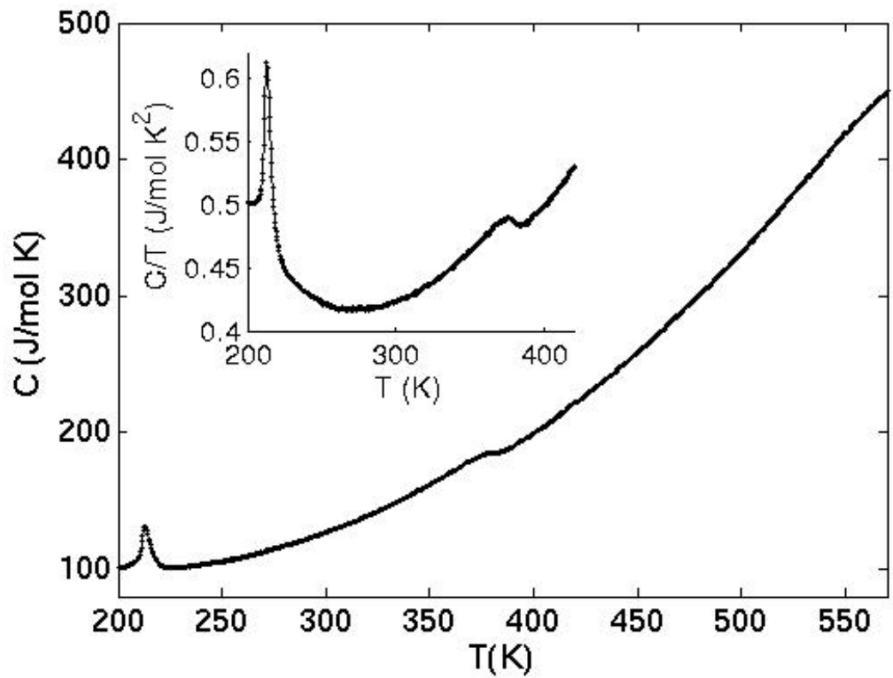

Figure 3

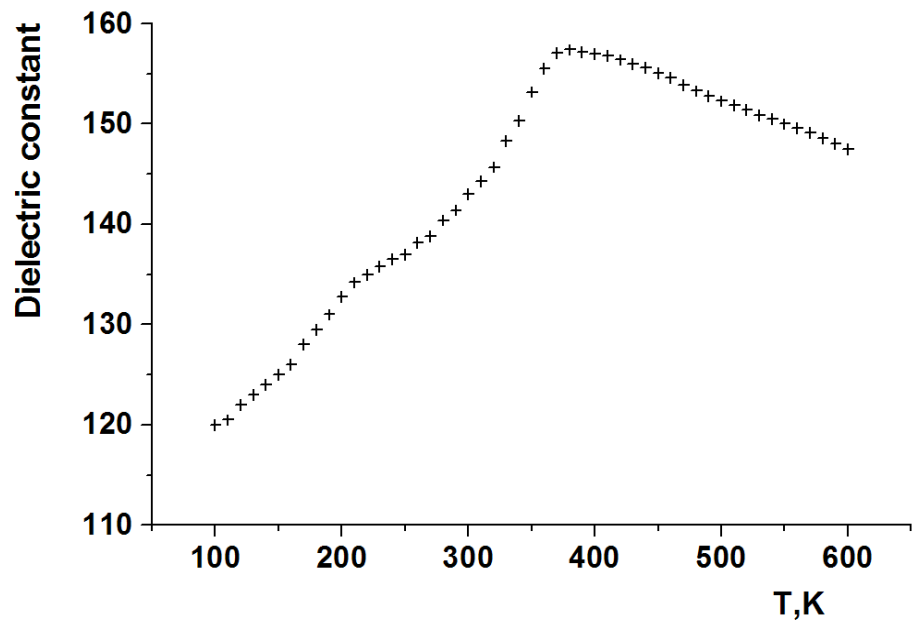

Figure 4

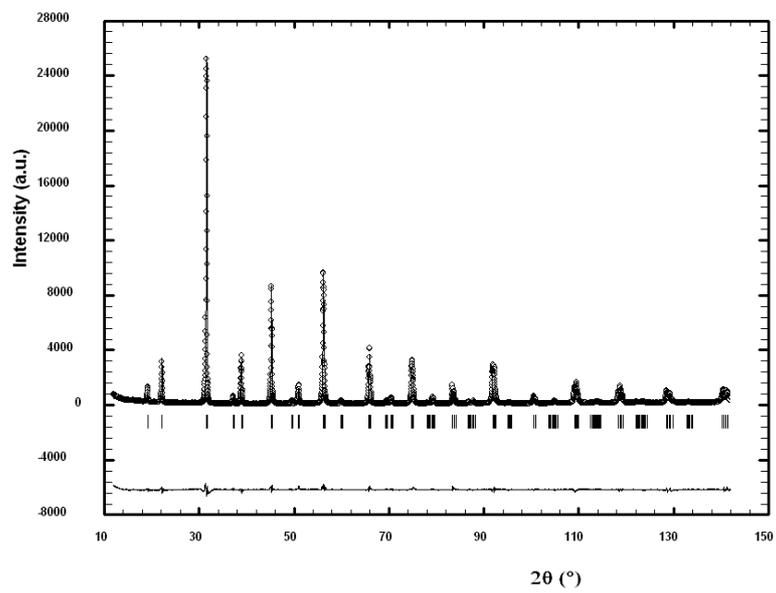

Figure 5

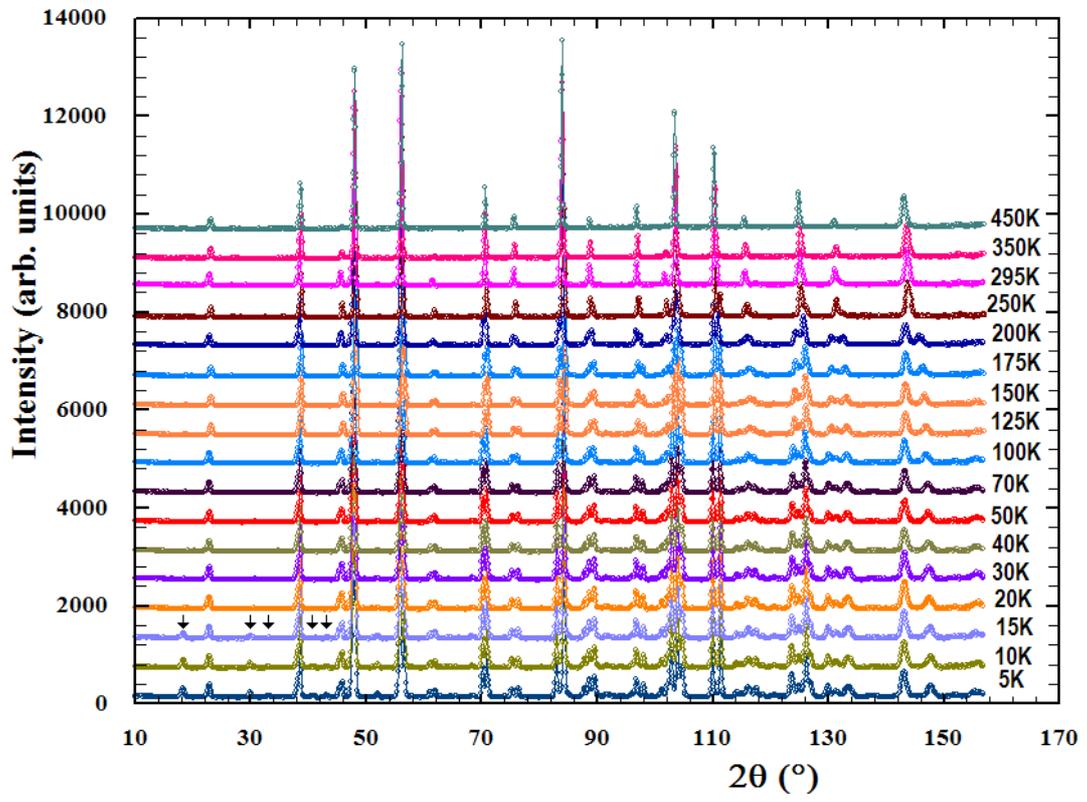

Figure 6

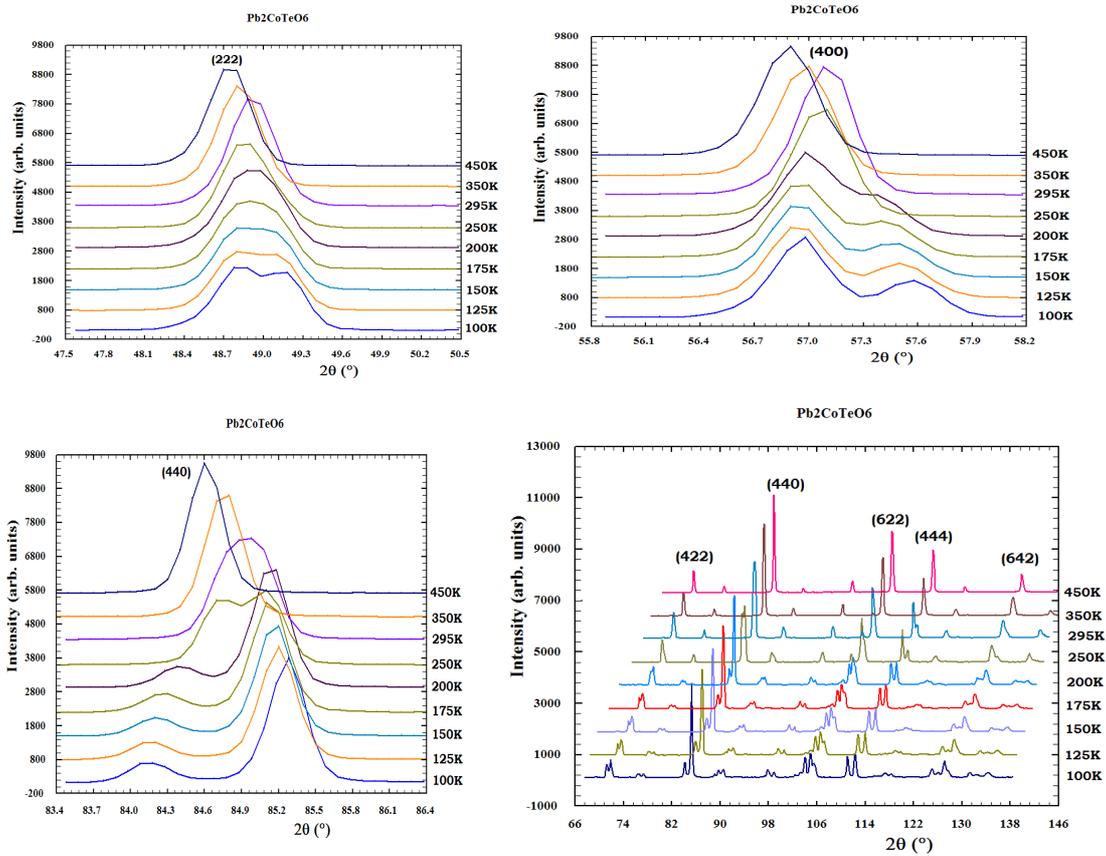

Figure 7

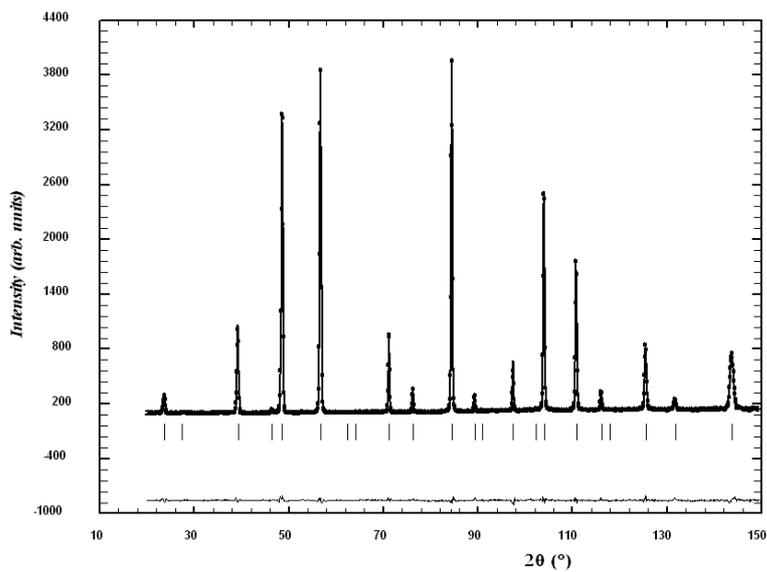

a)

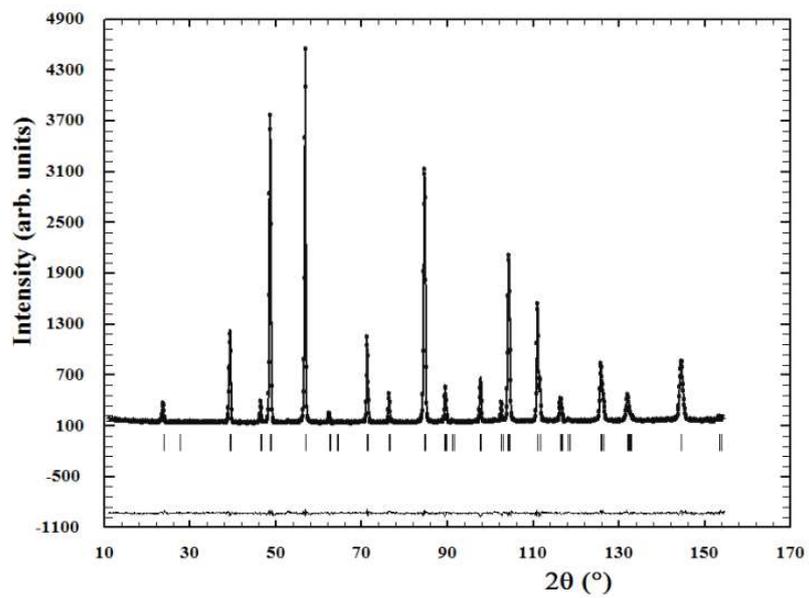

b)

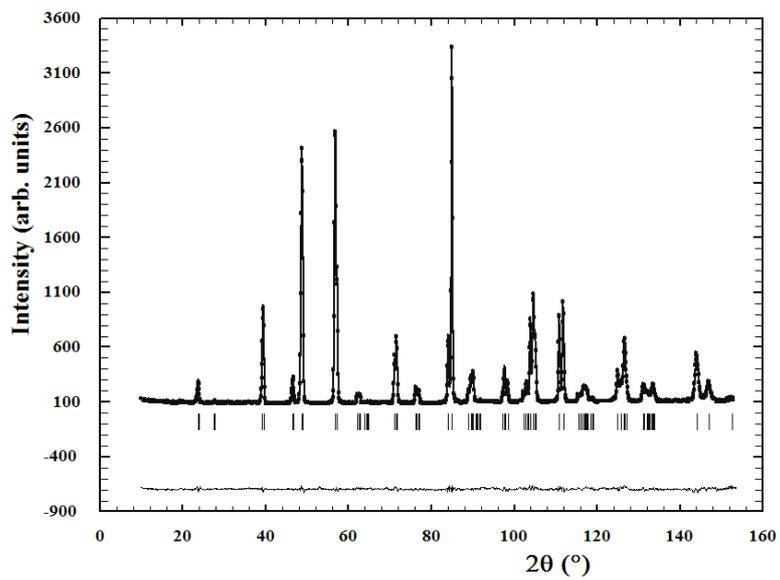

c)

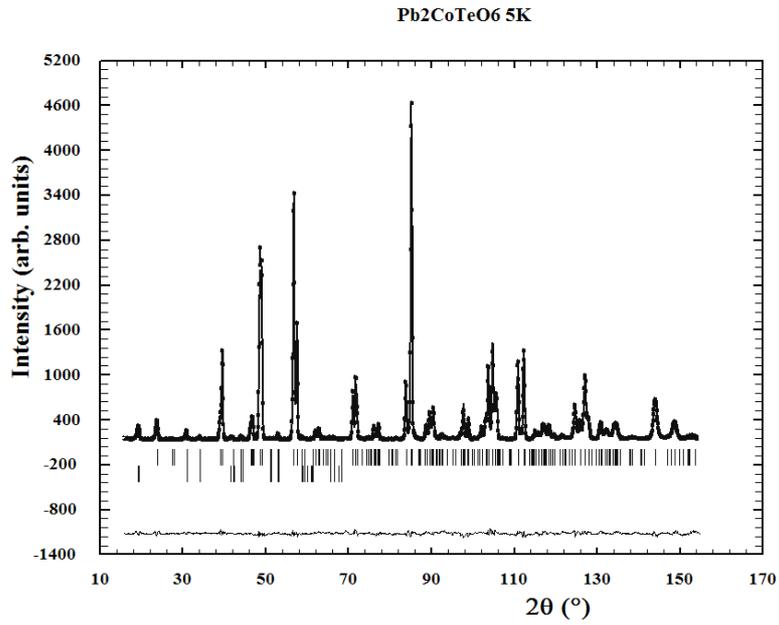

d)

Figure 8

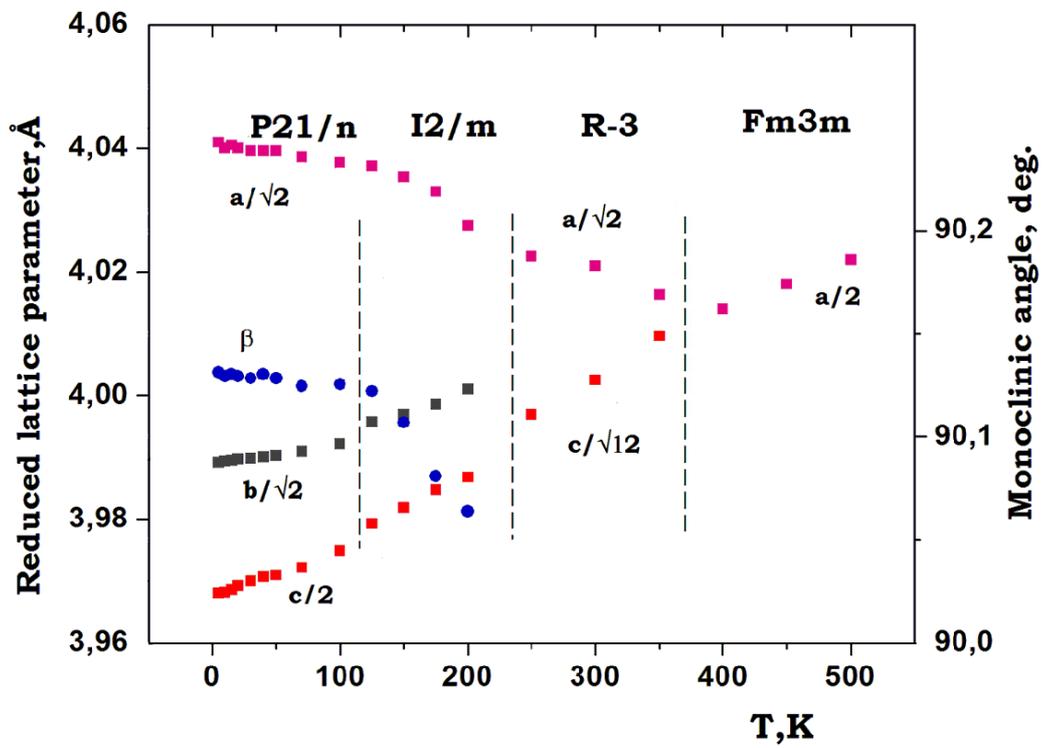

Figure 9

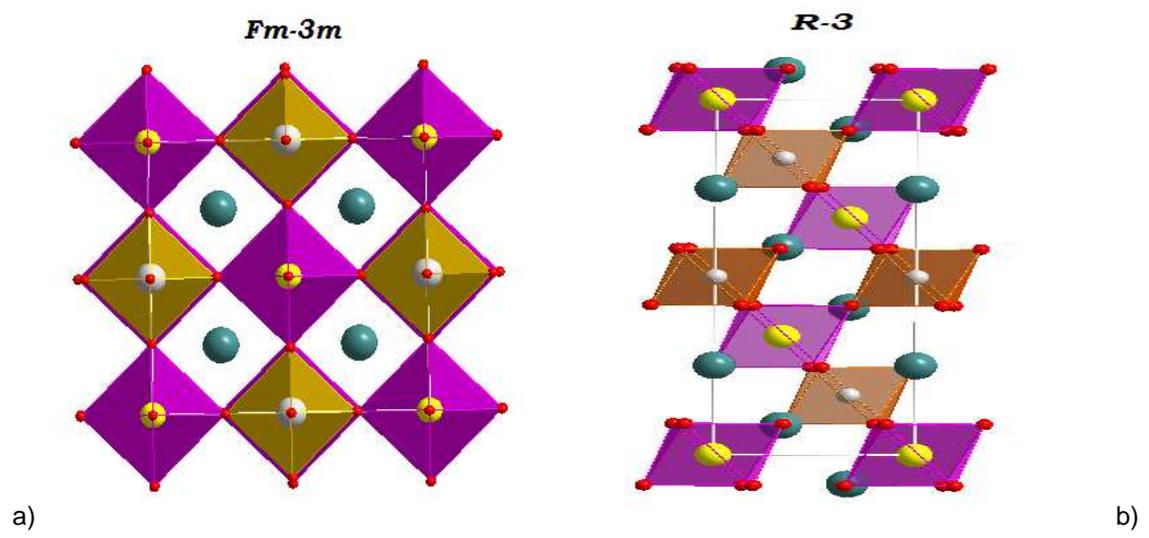

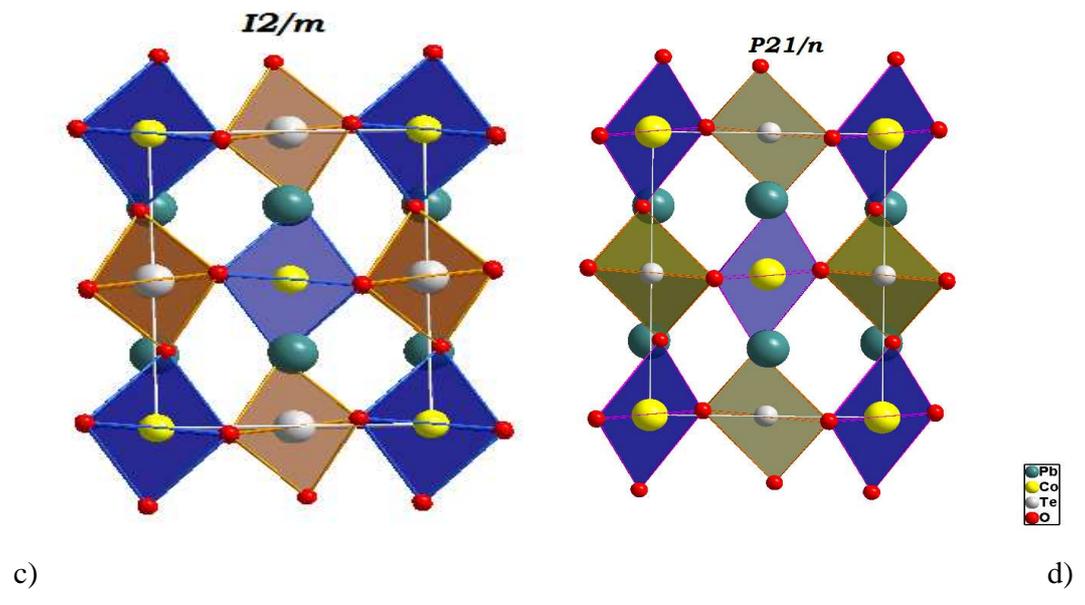

c)

d)

Figure 10

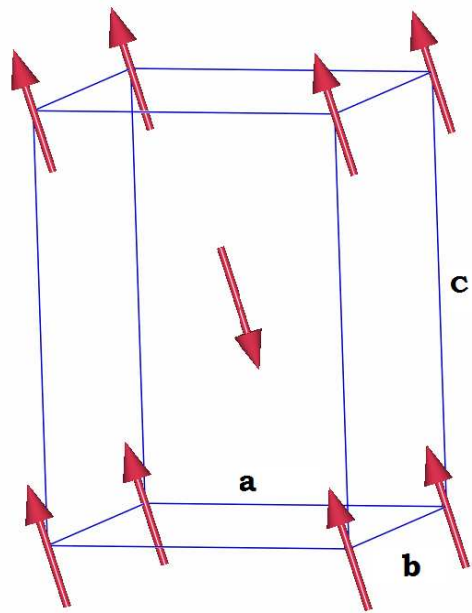

Figure 11